\documentclass{ws-procs9x6-cpt19}
\begin{document}

\newcommand{\refeq}[1]{(\ref{#1})}
\def\etal {{\it et al.}}
%any other macros go here 

\def\lrDnu{\stackrel{\leftrightarrow}{\! D}{\!}^{\lambda\,}}

\title{CPT- and Lorentz-Violation Tests with Muon $g-2$}

\author{B.\ Quinn}

\address{Department of Physics and Astronomy, University of Mississippi,\\
University, MS 38677, USA}

\author{On behalf of the Muon $g-2$ Collaboration}

\begin{abstract}
The status of Lorentz- and CPT-violation searches 
using measurements of the anomalous magnetic moment of the muon 
is reviewed. 
Results from muon $g-2$ experiments 
have set the majority of the most stringent limits 
on Standard-Model Extension Lorentz and CPT violation 
in the muon sector. 
These limits  are consistent 
with calculations of the level of Standard-Model Extension effects 
required to account for the current $3.7\sigma$ experiment--theory discrepancy 
in the muon's $g-2$. 
The prospects for the new Muon $g-2$ Experiment at Fermilab 
to improve upon these searches is presented. 
\end{abstract}

\bodymatter

\section{The anomalous magnetic moment}

\begin{figure}
\begin{center}
\includegraphics[width=3.5in]{{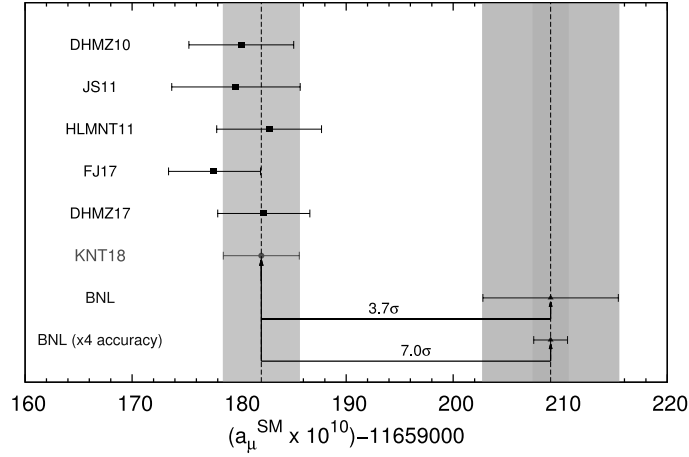}}
\end{center}
\caption{Comparison of SM evaluations of $a_\mu$ with the most recent experimental result and prospect.\cite{Keshavarzi}} 
\label{aba:fig1}
\end{figure}

The magnetic moment of the muon can be expressed by the relation
\begin{equation}
\vec{\mu} =
g\frac{e}{2m}\vec{s} =
(1 + a_{\mu})\frac{e}{m}\vec{s},
\label{aba:eq1}
\end{equation}
where the first term arises from the leading-order Dirac theory, 
and the anomaly, 
$a_{\mu} = (g - 2)/2$, 
represents the sum of all higher-order loop diagrams.\cite{Passera} 
The anomalous magnetic moment includes Standard-Model (SM) terms from QED, 
EW, 
and QCD processes, 
as well as possible contributions from Beyond the Standard Model (BSM) physics. 
The muon anomaly was measured to very high precision ($540\,$ppb) 
by the BNL E821 experiment
yielding $a^{\rm E821}_{\mu} = 116 592 089(63) \times 10^{-11}$.\cite{Bennett} 
When compared to the most recent SM calculations, 
the difference between the BNL result and theory is $3.7\sigma$, 
as shown in Fig.\ \ref{aba:fig1}.\cite{Keshavarzi} 
This discrepancy may be a sign of new physics. 
The new Muon $g-2$ experiment, 
E989, 
is currently running at Fermilab and aims to measure $a_\mu$ to $140\,$ppb, 
a factor of four improvement in precision.

\section{CPT- and Lorentz-violating signatures in $g-2$}

In the Muon $g-2$ experiment, 
a beam of polarized muons is injected into a storage ring. 
The anomaly is determined 
by measuring the ratio of two frequencies: 
$\vec{\omega}_a = \vec{\omega}_c - \vec{\omega}_s$, 
which is the rate at which the muon's spin ($\omega_s$) advances 
relative to its momentum ($\omega_c$), 
and the proton Larmor-precession frequency $\omega_p$, 
which is a measure of the ring magnetic field.\cite{TDR} 
These frequencies are related to the anomaly by
\begin{equation}
a_{\mu} =
\frac{\omega_a}{\omega_p}\frac{\mu_p}{\mu_e}\frac{m_\mu}{m_e}\frac{g_e}{2}.
\label{aba:eq2}
\end{equation}

The Standard-Model Extension (SME) is a general framework 
that describes CPT- and Lorentz-invariance violation 
by adding new terms to the SM lagrangian.\cite{Colladay} 
A minimal SME expression for the muon sector is
\begin{align}
\mathcal{L} = &
 - a_{\kappa}\bar{\psi}\gamma^{\kappa}\psi
 - b_{\kappa}\bar{\psi}\gamma_5\gamma^{\kappa}\psi
 - \tfrac{1}{2}H_{\kappa\lambda}\bar{\psi}\sigma^{\kappa\lambda}\psi %\label{aba:eq3}
\nonumber\\
& {}+ \tfrac{1}{2}ic_{\kappa\lambda}\bar{\psi}\gamma^{\kappa}\lrDnu\psi
 + \tfrac{1}{2}id_{\kappa\lambda}\bar{\psi}\gamma_5\gamma^{\kappa}\lrDnu\psi. \label{aba:eq4}
\end{align}
Equation \refeq{aba:eq4} predicts two Lorentz- and CPT-violating effects: 
a $\mu^+$/$\mu^-$ $\omega_a$ difference, 
$\Delta\omega_a = \langle\omega^{\mu^{+}}_a\rangle - \langle\omega^{\mu^{-}}_a\rangle$, 
and a sidereal $\omega_a$ variation.\cite{Bluhm}

In terms of the SME coefficients, 
$\Delta\omega_a = (4b_Z/\gamma)\cos{\chi}$, 
where $\chi$ is the colatitude of the experiment. 
Experimentally, 
it is convenient to perform the analysis on the ratio 
$\mathcal{R}=\omega_a/\omega_p$ from Eq.\ \refeq{aba:eq2}. 
The BNL E821 result $\Delta\mathcal{R} = -(3.6\pm3.7) \times 10^{-9}$ 
yields the limit $b_Z = -(1.0\pm1.1) \times 10^{-23}\,$GeV.\cite{BennettCPT} Comparison of $\omega^{\mu^{+}}_a$ from one experiment 
with $\omega^{\mu^{-}}_a$ from a second at a different colatitude 
affords sensitivity to the $d$ and $H$ coefficients, 
and doing so with BNL E821 and an earlier muon $g-2$ experiment at CERN\cite{CERN} 
gives $(m_{\mu}d_{Z0}+H_{XY}) = (1.6\pm5.6) \times 10^{-23}\,$GeV. 

Sidereal variation in $\omega_a$ 
is investigated using a Lomb--Scargle test\cite{LombScargle} 
for a significant amplitude of $\omega_a$ oscillation 
at the sidereal frequency. 
The Lomb--Scargle method is optimized for data unequally spaced in time, 
as is the case for E821. 
The limits on such an amplitude in the BNL data 
are $A^{{\mu}^-} < 4.2\,$ppm 
and $A^{{\mu}^+} < 2.2\,$ppm, 
which by the relationship $A^\mu = 2\check{b}^{\mu}_\perp\sin{\chi}$ 
is equivalent to $\check{b}^{{\mu}^-}_\perp \!\! \leq 2.6 \times 10^{-24}\,$GeV 
and $\check{b}^{{\mu}^+}_\perp \!\! \leq 1.4 \times 10^{-24}\,$GeV.\cite{BennettCPT} 
These BNL E821 limits
as well as other 
on both minimal ($d=4$) and nonminimal ($d \geq 5$) SME coefficients\cite{datatables} 
are the most stringent in the muon sector.

\section{Prospects for E989}
The goal for Muon $g-2$ at Fermilab 
is to reduce the BNL $a_\mu$ uncertainty 
by a factor of four from $540\,$ppb to $140\,$ppb. 
This will be achieved by utilizing Fermilab's much higher intensity muon beam 
to collect 21 times the BNL $\mu^+$ statistics, 
and by reducing the overall systematic uncertainty 
by a factor of 2.5 
through detector upgrades and improved analysis techniques. 
Reaching that goal would increase the significance of the BNL discrepancy 
from $3.7\sigma$ to $\sim7\sigma$ 
given the same central value for $a_\mu$.

With regard to sidereal-variation Lorentz and CPT tests, 
sensitivity roughly scales with $\omega_a$ uncertainty. 
Thus, 
E989 should be able to reach limits of $\sim 5 \times 10^{-25}\,$GeV
and could do even better 
due to the possibility to search for the oscillation 
with a Fourier-transform method 
since the E989 data will be time-stamped 
allowing binning in equally-spaced time periods. 
Also, 
the full three-year run for Muon $g-2$ 
will include data for most of the calendar year. 
Contrary to BNL E821,
which ran the same three months in each year of operation,
this permits 
a search for annual variation in $a_\mu$. 

Obviously, 
to measure $\Delta\omega_a$, 
Muon $g-2$ needs $\mu^+$ and $\mu^-$ data. 
The Fermilab Muon $g-2$ schedule 
features $\mu^+$ runs extending through early 2021. 
The Collaboration is exploring the technical requirements 
to carry out a $\mu^-$ run, 
as was done in E821. 
Items to be addressed include issues 
related to the lower initial muon flux 
and the need to improve the storage-ring vacuum. 
The optimal time for a switchover to $\mu^-$ 
depends on the results of the current, approved $\mu^+$ runs.

A $\mu^-$ run does not simply represent one more test. 
The $\mu^-$ data gives access to many additional SME coefficients. 
Furthermore, 
JPARC is preparing E34, 
a muon $g-2$ experiment with an ultra-cold muon beam.\cite{Otani} 
E34 proposes to measure $a_{\mu^+}$ to 450 ppb. 
This would make possible 
a substantial improvement of the $(m_{\mu}d_{Z0}+H_{XY})$ limit 
for two reasons. 
First, 
the BNL/CERN $540\,$ppm/$7000\,$ppm precisions 
would be replaced with Fermilab/JPARC $140\,$ppm/$450\,$ppm. 
Second, 
this limit is proportional to $(\cos{\chi_1}-\cos{\chi_2})$, 
and there is much greater difference 
between Fermilab's and JPARC's colatitudes 
then between BNL and CERN. 
Because E34 utilizes muonium, 
and thus cannot measure $a_{\mu^-}$, 
the only possibility for realizing this potential improvement 
is with a Fermilab Muon $g-2$ $\mu^-$ run.  

Finally, 
it has been shown 
that a nonminimal SME coefficient 
$\check{H}^{(5)}_{230} \simeq 3 \times 10^{-25}\,$GeV$^{-1}$
can account for the $3.7\sigma$ discrepancy in muon $g-2$.\cite{Gomes} 
%For example, 
%$\check{H}^{(5)}_{230} \simeq 3 \times 10^{-25}\,$GeV$^{-1}$ 
%would be sufficient to produce the measured difference in $a_\mu$. 
The result from BNL E821 of $\check{H}^{(5)}_{230} = (2.9 \pm 3.0) \times 10^{-24}\,$GeV$^{-1}$ is compatible with this level. 
With an E989 $\mu^-$ run and the promise of Fermilab and JPARC sensitivity goals, 
it may be possible to not only establish a significant discrepancy, 
but also make a statement concerning whether or not Lorentz and CPT violation 
is the physics responsible for it.

%Finally, 
%calulations were performed 
%to determine the size of $d=5$ $H$ coefficients 
%required to account for the muon $g-2$ $3.7\sigma$ discrepancy.\cite{Gomes} 
%For example, 
%$\check{H}^{(5)}_{230} \simeq 3 \times 10^{-25}\,$GeV$^{-1}$ 
%would be sufficient to produce the measured difference in $a_\mu$. 
%The result from BNL E821 of $\check{H}^{(5)}_{230} = (2.9 \pm 3.0) \times 10^{-24}\,$GeV$^{-1}$ is compatible with this level. 
%With an E989 $\mu^-$ run and the promise of Fermilab and JPARC sensitivity goals, 
%it may be possible to not only establish a significant discrepancy, 
%but also make a statement concerning whether or not Lorentz and CPT violation 
%is the physics responsible for it.

\section*{Acknowledgments}
This work was supported in part by the US DOE and Fermilab 
under contract no.\ DE-SC0012391. 
The author is grateful to Alan Kosteleck\'y 
for the CPT'19 invitation and conversations.

\end{document}